# Scientific Exploration with Expert Knowledge (SEEK) in Autonomous Scanning Probe Microscopy with Active Learning

Utkarsh Pratiush[1], Hiroshi Funakubo[2], Rama Vasudevan[3], Sergei V. Kalinin[1,4**], Yongtao Liu[3*]

1. Department of Materials Science and Engineering, The University of Tennessee Knoxville, Knoxville, Tennessee 37996, United States of America
2. Department of Materials Science and Engineering, Tokyo Institute of Technology, Yokohama, 226-8502, Japan
3. Center for Nanophase Materials Sciences, Oak Ridge National Laboratory, Oak Ridge, TN, 37830, United States of America
4. Physical Sciences Division, Pacific Northwest National Laboratory, Richland, Washington 99354, United States of America

Corresponding emails:
    \*   liuy3@ornl.gov
    \**  sergei2@utk.edu

**Abstract**

Microscopy techniques have played vital roles in materials science, biology, and nanotechnology, offering high-resolution imaging and detailed insights into properties at nanoscale and atomic level. The automation of microscopy experiments, in combination with machine learning approaches, is a transformative advancement, offering increased efficiency, reproducibility, and the capability to perform complex experiments. Our previous work on autonomous experimentation with scanning probe microscopy (SPM) demonstrated an active learning framework using deep kernel learning (DKL) for structure-property relationship discovery. This approach has demonstrated broad applications in various microscopy techniques. Here, to address limitations of workflows based on DKL, we developed methods to incorporate prior knowledge and human interest into DKL-based workflows and implemented these workflows in SPM. By integrating expected rewards from structure libraries or spectroscopic features, we enhanced the exploration efficiency of autonomous microscopy, demonstrating more efficient and targeted exploration in autonomous microscopy. We demonstrated the application of these methods in SPM, but we suggest that these methods can be seamlessly applied to other microscopy and imaging techniques. Furthermore, the concept can be adapted for general Bayesian optimization in material discovery across a broad range of autonomous experimental fields.

**Introduction**

Microscopy techniques are indispensable tools in materials science, biology, and nanotechnology, driving innovations by providing insights into the structural and functional characteristics of materials at the micro, nano, and atomic scales. Techniques such as Atomic Force Microscopy (AFM) have revolutionized our ability to visualize and manipulate matter with nanoscale precision[1]. AFM not only offers high-resolution imaging by scanning a sharp probe over sample surface for investigation of mechanical[2,3], electrical[4–6], and chemical properties[7,8] at the nanoscale; but also, AFM spectroscopy offers detailed insights into local dynamics by applying temporal excitation at nanoscale structures[9–15].

The automation of microscopy experiments is one of the most transformative advancements in the field[16–25]. Automated microscopy can offer comprehensive insights into materials properties, systematically studying materials behavior under a spectrum of conditions by adjusting experiment parameters such as scan size and excitation bias in an automatic manner[26–28]. This significantly enhances efficiency and precision of microscopy experiment. These automated systems bring numerous benefits, including increased throughput, improved reproducibility, and the ability to perform experiments that would be impractical or impossible manually.

The rapid rise of machine learning (ML) applications in microscopy[29–36] has also significantly enhanced the field by assisting instrument tuning[37,38], data processing and analysis[39–45], and acquisition[46–50]. In microscopy data analysis, ML can process vast amounts of data to uncover patterns and correlations that are not immediately apparent to human analysts[51–57], accelerating the analysis process. In on-the-fly experiment, ML facilitates the development of autonomous experimentation systems by leveraging microscopy automation, ML algorithms, and workflow integration to perform experiments with minimal human intervention[23,58,59]. By integrating ML, researchers can design experiments, optimize imaging conditions, and even make decisions about subsequent measurements based on real-time data analysis[47,50,60]. This level of automation and intelligence is transforming microscopy into a more powerful and efficient tool, driving advancements in fields such as materials science, nanotechnology, and biology.

Generally, by now three levels of the ML applications in microscopy can be defined. On the simplest level, ML is used as a part of data analysis after the experiment. These applications offer a broader toolbox of image and analysis method compared to classical image analysis tools, but do not significantly change the nature of microscope experiment[33,38,61–66]. The implementation of the ML image analysis methods as a part of experiment offers real time segmentation and dimensionality reduction of data[67–69], ultimately facilitating representation of high dimensional and complex data to human operator. This significantly increases requirements placed on to the used ML algorithms, particularly towards out of distribution shifts[67]. However, real-time image analytics still requires human decision making. Finally, the third level of ML implementation is having ML agents directly controlling the instrument though suitable API or software library. In this approach, the decision can be purely ML diven[43,70,71], or the behavior of the ML agent including decision making policies and reward functions can be continuously tuned by human operator, giving rise to human in the loop approach[19,33,34,61,72].

Previously we have implemented the autonomous experiment (AE) in scanning probe microscopy (SPM) for structure-property discovery using the static and dynamic policies. In the former, the pretrained ML algorithms identifies a priori know object of interest and performs specific experiments using predefined policies. The example of this approach is to identify domain walls in ferroelectric samples and grain boundaries in hybrid perovskites[12,46], enabling the discoveries of high responsive ferroelastic domain walls and insulating grain boundary junctions, respectively.

The alternative is the spectral discovery experiments, where an ML algorithm aims to discover which microstructures optimize the certain spectral features. Unlike the static policy experiment where ML performs human-level semantic segmentation task, this is an example of beyond-human AE. We recently implemented an active learning framework using deep kernel learning (DKL)[50]. DKL integrates a neural network with a Gaussian process (GP) layer. Specifically, the neural network transforms 2D structural images into a low dimensional latent space, while scalarizer functions, based on prior knowledge or expert expectations, convert spectroscopic data into scalar physical descriptors. Practically, the scalarizer serves as a reward function defining the degree of experimentalist' interest towards specific spectral features. GP then operates over the latent space to uncover the relationship between image and spectroscopic data, namely the structure-property relationship. This approach has demonstrated broad applications in atomic force microscopy[22], scanning tunneling microscopy[73], and scanning electron transmission microscopy[33,47], which can also be adapted for any other imaging and spectroscopy techniques. DKL enhances the capabilities of these systems by providing real-time data analysis and decision-making, enabling the identification of image patterns that are potentially interesting for spectroscopic investigation. In the process, DKL equally explore the entire region; however, in materials science, intriguing physics or functionality is often associated with specific structures or embedded in particular spectroscopic features. Instead of equally exploring the entire structure library and spectroscopic results, prior knowledge and expertise can be applied to refine the exploration space in DKL to enhance the discovery of intriguing functionality.

Both the static and simple dynamic workflows have a number of limitations with respect to broad variability of surface responses and tunability of reward functions. Here, we report on methods to incorporate prior knowledge and human interest into DKL-driven microscopy, engendering the transition from simple discovery loops to the multi-stage decision making. We refer to this approach as Scientific Exploration with Expert Knowledge (SEEK). We demonstrate that the prior knowledge and expected reward can be integrated both from the structure library or the spectroscopic features for DKL exploration. When the constraint is applied to the structure, a machine learning method can analyze the acquired full structural image, identify the structure of interest based on human knowledge, and then form a structure library that includes only the structures of interest for DKL exploration. When the constraint is applied to the spectroscopic features, a second surrogate model predicts the target spectroscopic features, interacting with the acquisition function to determine the next measurement location. We implemented these approaches in both pre-acquired AFM model data and operating AFM experiment, revealing a more efficient exploration of autonomous microscopy.

## *Traditional DKL-driven discovery*

Figure 1 illustrates the process of a standard DKL-driven exploration, which begins with acquiring a 2D structural image, such as topography or PFM images. From this structural image, we create image patches centered on individual pixels, which form a structure library representing the nanoscale spatial structure. Detailed properties at each pixel are measured via a spectroscopic mode, such as piezoresponse spectroscopy, force-displacement spectroscopy, current-voltage curve, etc. The acquired spectrum is analyzed by a pre-defined scalarizer function to extract a physical descriptor for DKL training. At the beginning of a DKL-driven experiment, a few spectra are acquired at random or strategically selected locations. The DKL model is then trained using structure patches from these locations and corresponding physical descriptors from the measured spectra. The DKL model learns a probabilistic relationship between structure patches and physical descriptors, predicting the physical descriptor values for structure patches where spectra have not yet been measured. Utilizing DKL prediction and uncertainty for unmeasured pixels, an acquisition function determines the next structure patch (pixel) for spectroscopic measurement. After each new measurement, the DKL model is retrained with the updated training dataset, incorporating new structure patches and physical descriptors, thereby iterating the process. This iterative approach enhances the efficiency and accuracy of discovering structure–property relationships in high-dimensional datasets.

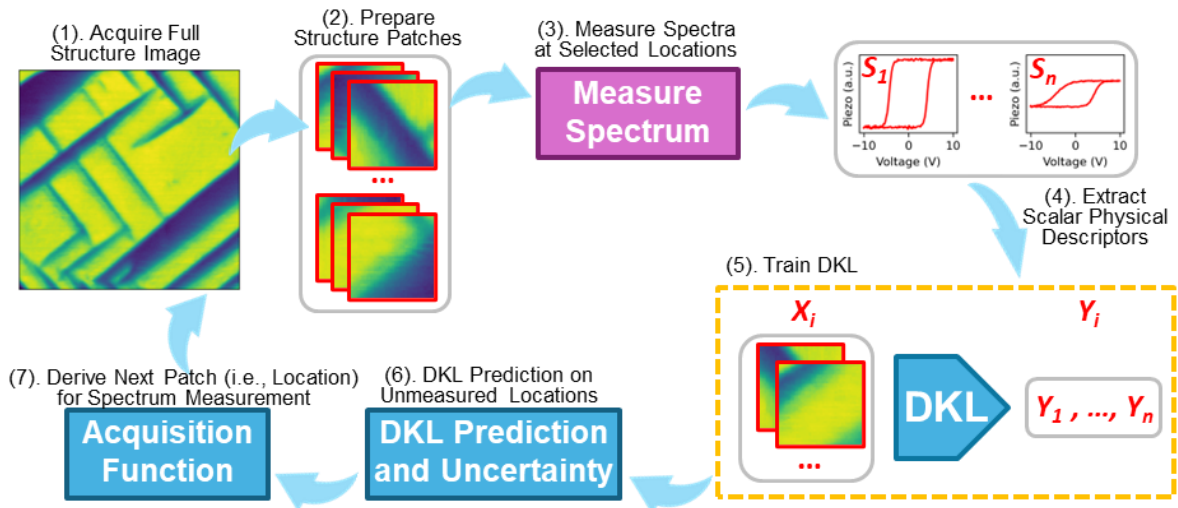

*Figure 1. Traditional DKL driven discovery. Image patches representing local structures are prepared on dense grid locations within an acquired full structure image. At the beginning, a few spectra can be measured at random locations, all spectra are analyzed by a scalarizer function to convert 1D spectra data to scalar physical descriptors. The corresponding image patches and descriptors at measured locations form a training dataset for DKL training, followed by DKL prediction at unmeasured locations. Then, an acquisition function derives the next spectrum measurement location and microscopy performs the next measurement, the process is repeated until a certain criterion is achieved.*

*Structure-constrained DKL*

Typically, structure patches are prepared at uniform dense grid locations within a full structure image (as shown in Figure 1), as such DKL equally explores the entire region under study, the choice of spectroscopic measurement location is determined solely by the acquisition function. However, in materials research, properties of interest are often concentrated in specific locations, such as domain walls in ferroelectric materials, or grain boundaries in photovoltaic perovskites. Focusing on structure patches related to these specific locations could further accelerate discoveries. To leverage this prior knowledge, we can constrain the preparation of structure patches to center only on the pixels that correspond to the structures of interest, then resultant structure library will exclusively represent structures of interest (Figure 2). Consequently, the DKL exploration will be focused solely on these structures of interest, enhancing the efficiency of discovering structure-property relationships in these critical areas.

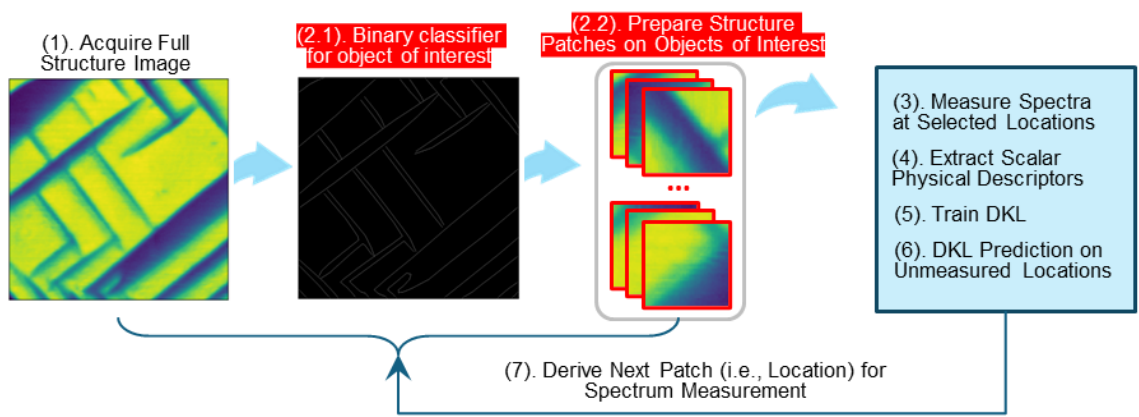

*Figure 2. Structure-constrained DKL. Before preparing the structure patches, the structure of interesting (e.g., ferroelastic domain walls) can be identified from the full structure image using a neural network or other method (i.e. 2.1.), then the structure patches are only prepared on these identified locations to form a structure library (i.e. 2.2) only containing structures of interest for DKL exploration. As such, the DKL-driven discovery will specifically focus on these structures of interest. In the workflow, the neural network (2.1) either serves as a binary classifier that defines objects as interesting/not interesting or ascribes the "interest score" to each patch that can be used to sample from when choosing next object. This neural network can be trained, or the method can be defined prior to the experiment based on human's interest and knowledge.*

We first showcase the application of the structure-constrained DKL in a model dataset. This model dataset is a vertical band excitation piezoresponse spectroscopy (BEPS) data of a PbTiO3 (PTO) thin film. This PTO material has been explored via various machine learning empowered automated and autonomous microscopy previously[22,46,50,74], suggesting that it is a good model system for developing ML in materials science. Figure 3a shows a PFM amplitude image from the BEPS hyperspectral dataset, displaying both in-plane *a*-domains (dark color) and out-of-plane *c*-domains (green color). This amplitude image consist of detailed nanoscale domain structures will be the full structure image for DKL exploration. Figure 3b shows a few representative piezoresponse vs. voltage hysteresis loops from marked locations in Figure 3a of the structure image.

The polarization vector in *a*-domains is parallel to the sample surface, making it insensitive to vertical BEPS measurements that characterize out-of-plane polarization dynamics, resulting in closed hysteresis loops. In contrast, *c*-domains have a polarization vector perpendicular to the sample surface, resulting in open hysteresis loops that reveal in-depth ferroelectric characteristics (e.g., polarization magnitude, coercive field, nucleation bias, etc). Therefore, when exploring the relationship between ferroelectric domain structures and hysteresis loops using vertical BEPS, it is more effective to sample hysteresis loop measurement locations at *c*-domains with open hysteresis loops for detailed ferroelectric characteristics. To achieve this, a mask (e.g., threshold filter) can be applied to the full structure image to identify *c*-domains. For example, Figure 3c shows *c*-domains identified by a **binary** threshold filter, where the yellow regions represent c-domains. When a structure library is created from locations corresponding to *c*-domains, DKL-driven discovery will focus on exploring these *c*-domains. This is demonstrated by the acquisition map shown in Figures 3d, which only samples effective acquisition values at *c*-domains. As a comparison, the acquisition map of a traditional DKL is also plotted in Figure 3e, which samples acquisition at all unmeasured locations. Notably, traditional DKL results in higher acquisition values at *a*-domains, leading to more spectroscopy measurements at *a*-domains that are insensitive to vertical BEPS measurements, leading to inefficient exploration. The detailed comparison in the performance of structure-constrained DKL and traditional DKL will be discussed later.

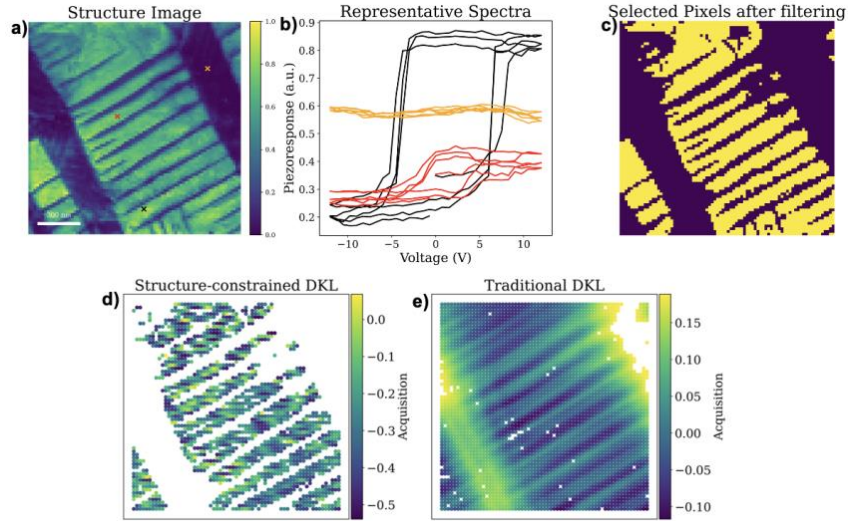

*Figure 3. Structure-constrained DKL. (a) The full structural image is PFM amplitude that showing a- and c-domains, the scale bar is 300 nm. (b) a few representative piezoresponse vs. voltage hysteresis loops from the locations marked in (a). (c) before preparing structure patches, a threshold filter is used to extract the c-domain locations that is shown as yellow, c-domains are expected to result in more meaningful hysteresis loops consist of in-depth ferroelectric characteristics. (d) the DKL acquisition map after 100 exploration steps, which only contains effective samplings at c-domains. (e) in contrast, traditional DKL acquisition map indicates higher acquisition values at a-domains that is insensitive to the vertical BEPS measurement, leading to ineffective exploration steps.*

### *Spectrum-constrained DKL*

However, very often the prior knowledge is associated with the spectral rather than structural behavior, and is not available prior to the experiment except in the most general form. Here, we demonstrate how the prior knowledge can also be applied to the real time spectroscopic data to guide DKL driven discovery as a two-step decision process.

Here, we note that the scalarizer function, which converts spectra into physical descriptors, is typically predefined and universally applied to each spectrum throughout the experiment. However, some spectra may not be properly processed by this predefined scalarizer function. This issue does not simply arise from noisy raw spectra but could also indicate a shift to unknown physics. For example, if a scalarizer function is designed to find the peak position, assuming only one peak appears in the spectrum, two problematic scenarios might occur: First, some spectra may show no peak, likely due to noise. This scenario can be anticipated prior to the experiment, and proper design of the scalarizer function may help handle it. Second, some spectra may reveal multiple peaks or even inverted peaks, which are more complex and unpredictable before the experiment. The predefined scalarizer function cannot effectively process this type of raw spectrum, resulting in outlier of scalar descriptors and potentially contaminating the DKL training dataset. This second scenario may indicate new physics.

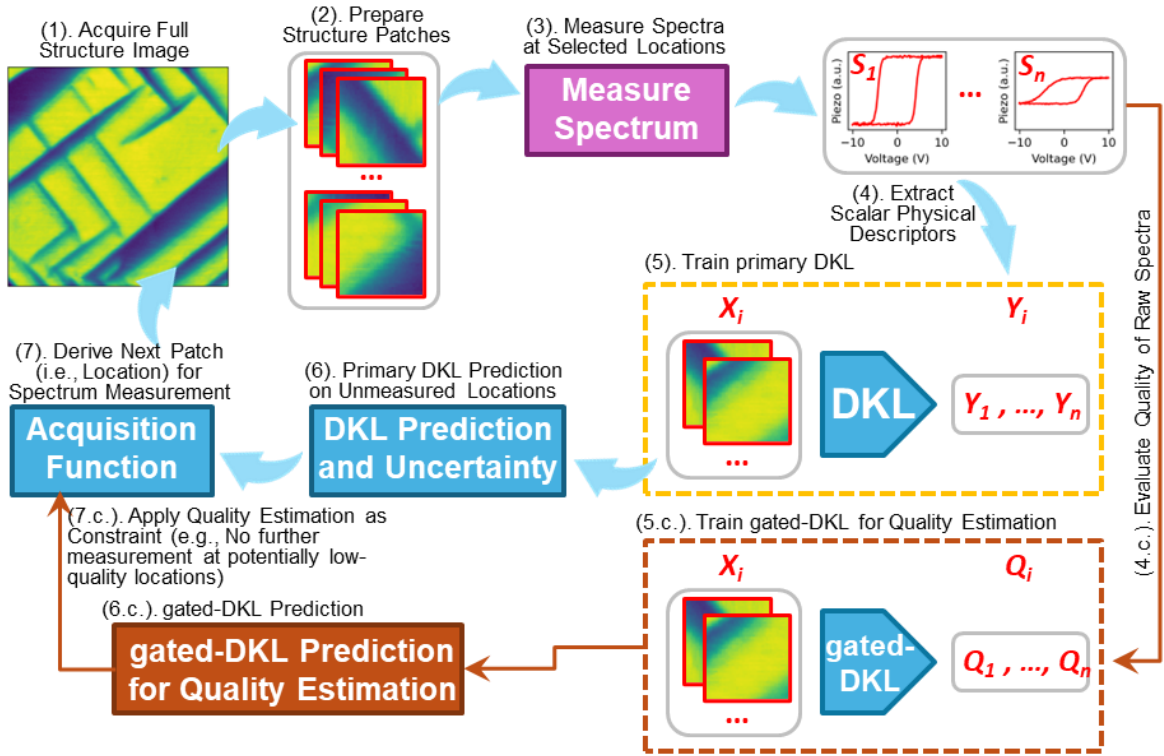

Figure 4. Spectrum-constrained DKL, in addition to the primary DKL for driving discovery, a gated-DKL is used to predict spectra quality of unmeasured locations, such prediction can reflect the occurrence of unusual raw spectra that cannot be properly process by the pre-defined scalarizer function. Then this gated-DKL prediction can be used to adjust the acquisition map of the primary DKL accordingly, ensuring the primary DKL sampling focuses on the structure space possibly resulting in spectra that can be properly processed by the predefined scalarizer function. Note that here the spectra quality is a metric representing whether the spectra can be properly processed by the predefined scalarizer function, in some cases, a low-quality score can possible because of the appearance of new physics that are not captured in the scalarizer function, which is defined based on human prior knowledge.

To address the aforementioned challenges, we develop two-stage decision making process. Here, in addition to the primary DKL, a gated-DKL is used to evaluate the real-time raw spectrum and predict the occurrence of unusual spectra data, as shown in Figure 4. By using the gated-DKL to predict where in the structural space these unusual spectra might appear and adjusting the acquisition map of the primary DKL accordingly, the primary DKL can focus on the structural space possibly resulting in more relevant spectrum data, namely, spectrum constrained DKL. With the assistance of gated-DKL, the primary DKL has the potential to accelerate the discovery of the target property in the on-the-fly experiment; in addition, the gated-DKL can highlight the structural space with high probability of unusual spectra data, implicating the need for deeper analysis of the spectra from the highlighted space to uncover new physics.

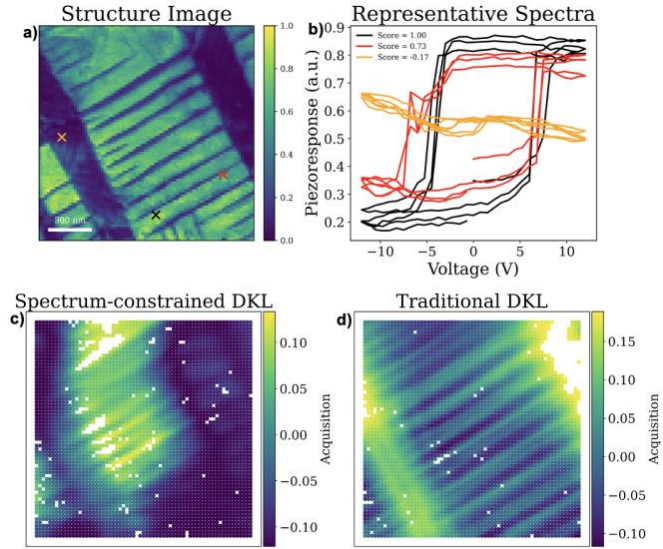

*Figure 5. Spectrum-constrained DKL. (a) the full structure image is PFM amplitude, the scale bar is 300 nm. (b) Several representative spectra with their respective quality score, here the quality score is obtained by calculating the structure similarity index between respective spectrum and a reference spectrum. We selected a hysteresis loop (black one) with large loop opening as the reference spectrum, as opening loop can offer more ferroelectric characteristics than a closed loop. (c) the acquisition map of spectrum constrained DKL showing larger acquisition value of the c-domains that are responsive to vertical BEPS measurement. (d) the acquisition map of a traditional DKL showing larger acquisition value of a-domains that are insensitive to vertical BEPS measurement.*

The spectrum-constrained DKL approach, utilizing two DKLs, is implemented on the pre-acquired PTO BEPS dataset to test its performance. Figure 5a shows the full structural image used for the spectrum-constrained DKL exploration. In this approach, the gated-DKL evaluates the quality of the acquired raw spectra and predicts the spectra quality at all unmeasured locations. To evaluate the quality of the acquired spectra, we employed the structural similarity index as a quality metric, calculating the structural similarity index between the acquired spectra and a reference spectrum. The reference spectrum is a representative of high-quality spectrum data, which can be selected from the seed spectra measurements or defined by researchers. In this study, we used a spectrum from the seed measurement as the reference spectrum (black spectrum in Figure 5b). Figure 5b shows the quality scores of a few representative spectra calculated using this approach.

The gated-DKL is trained on the quality of the acquired raw spectra and predicts the quality scores at unmeasured locations. These predicted quality scores are then used to adjust the acquisition map of the primary DKL, ensuring that the primary DKL-driven discovery focuses on areas likely to yield high-quality raw spectra data. Figure 5c shows the acquisition map of the spectrum-constrained DKL, indicating high acquisition in *c*-domains. This is because *c*-domains are more responsive to vertical BEPS measurements, resulting in opening hysteresis loops similar to the reference spectrum. In contrast, the traditional DKL acquisition (Figure 5d) results in more sampling in *a*-domains, which are insensitive to vertical BEPS measurements.

**Comparing explorations of traditional DKL and constrained DKLs**

Next, we compare the explorations of traditional DKL, structure-constrained DKL, and spectrum-constrained DKL. In these DKL explorations, the hysteresis loop area is used as the reward target, which indicates hysteresis loop opening. Opening hysteresis loops provide more detailed information regarding ferroelectric characteristics, such as remnant polarization, coercive field, and nucleation bias. Therefore, explorations that acquire more opening hysteresis loops can potentially be more effective. Figure 6a plots the hysteresis loop area as a function of exploration steps. It is evident that the traditional DKL sampled many closed hysteresis loops (loop areas close to zero). In contrast, both spectrum-constrained DKL and structure-constrained DKL sampled more opening hysteresis loops (loop areas deviating from zero). The sampling locations are shown in Figures 6b-d. Traditional DKL focused on *a*-domains, which are insensitive to vertical BEPS measurements, leading to ineffective measurements. However, spectrum-constrained DKL and structure-constrained DKL targeted *c*-domains that are more responsive to the vertical BEPS measurement. The loop area distributions of the three DKL approaches are shown as histograms in Figures 6e-g. The constrained DKL methods (Figures 6f-g) sampled more opening loops with meaningful ferroelectric characteristics than the traditional DKL (Figure 6e), indicating a more effective exploration of constrained DKL methods that incorporate prior knowledge as constraints. Specifically, the pre-acquired data indicate that the percentage of sampled opening loops with meaningful ferroelectric characteristics (loop area more than 0.2) is 3% for the traditional DKL method, 10% for the spectrum-constrained DKL method, and 34% for the structure-constrained DKL method.

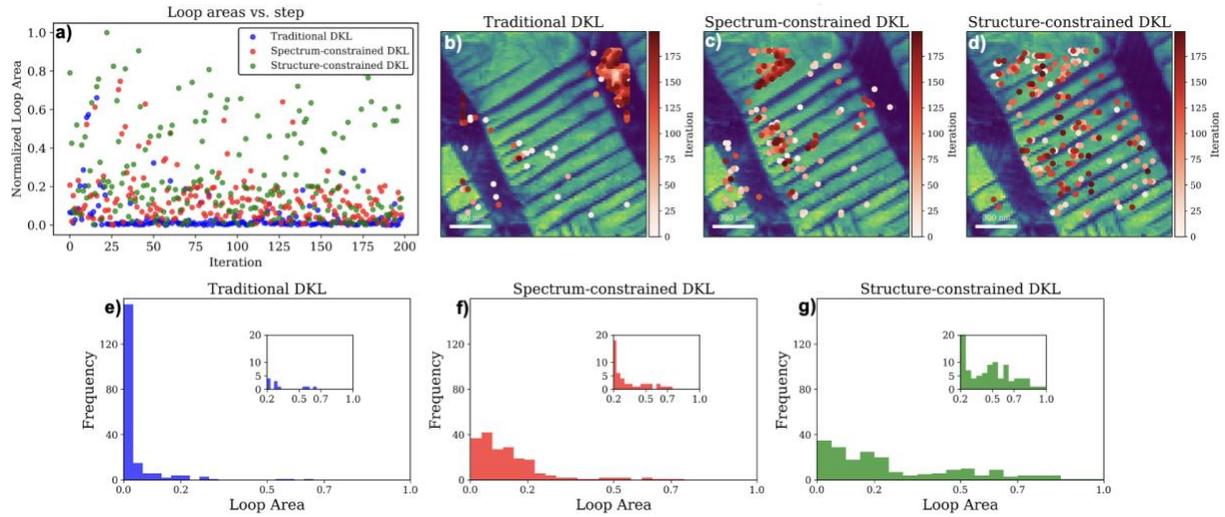

*Figure 6. Comparison of traditional DKL, structure-constrained DKL, and spectrum-constrained DKL. (a) Hysteresis loop area as a function of DKL exploration steps. (b)-(d) Sampled measurement locations of traditional DKL, spectrum-constrained DKL, and structure-constrained DKL, respectively. (e-g) histogram of loop areas of traditional DKL, spectrum-constrained DKL, and structure-constrained DKL, respectively.*

### *Implementing in operating autonomous microscopy*

After analyzing the performance of constrained DKL using pre-acquired data with known ground truth, we implemented these approaches in operating microscopy for autonomous experiments. Specifically, we applied these methods in band excitation piezoresponse force microscopy (BEPFM) and spectroscopy (BEPS) using our AEcroscopy platform[18] to investigate the ferroelectric PTO thin film. AEcroscopy is a platform for automated experiments in scanning probe and electron microscopy, allowing seamlessly implementation of machine learning approaches in operating microscopy for accelerating discoveries, details about AEcroscopy can be found in our previous report[18].

Compared to the pre-acquired hyperspectral BEPS data, the operating PFM can provide high spatial resolution BEPFM images that clearly illustrated domain walls. This allowed us to apply a structural constraint of domain walls in addition to the previously mentioned structural constraint of *c*-domains. To implement the structural constraint of domain walls, we utilized a previously developed ensembled neural network for real-time identification of ferroelastic domain walls. This method identifies domain wall locations from on-the-fly PFM images. Using these identified locations, we prepared structural image patches focused specifically on the extracted domain walls, enabling domain wall constrained DKL exploration.

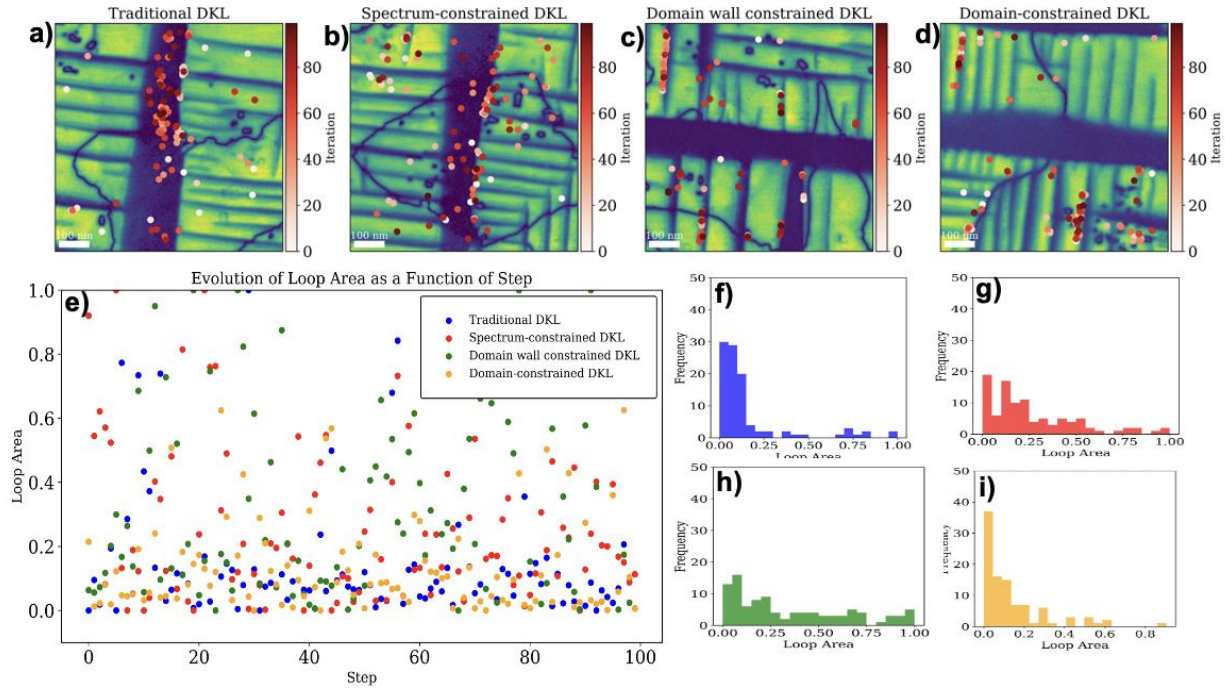

Figure 7. Constrained DKL in Operating Autonomous Microscopy. (a)-(d) Sampled BEPS measurement locations of traditional DKL, spectrum constrained DKL, domain wall constrained DKL, and domain constrained DKL, respectively. (e) hysteresis loop area as a function of DKL exploration steps. (f-i) histograms of loop area distribution of traditional DKL, spectrum constrained DKL, domain wall constrained DKL, and domain constrained DKL, respectively.

Figure 7 presents the results of DKL explorations in operating autonomous PFM, comparing traditional DKL, spectrum-constrained DKL, domain wall-constrained DKL, and domain-constrained DKL. Figures 7a-d shows the DKL sampling locations on the full structural image of the PFM amplitude. It is seen that traditional DKL predominantly samples measurement locations in *a*-domains, which are insensitive to vertical BEPS measurements, leading to ineffective measurements. In contrast, when constraints based on prior knowledge are applied (Figures 7b-d), either spectrum-constrained or structure-constrained, the constrained DKL directs measurements to potentially intriguing locations. For example, spectrum-constrained DKL (Figure 7b) primarily samples BEPS measurements near a-c domain walls. Similarly, domain-constrained

DKL samples measurements at either a-c domain walls or c-c domain walls, which are known to encode intriguing properties in ferroelectrics.

Figure 7e plots the hysteresis loop areas as a function of exploration steps, with corresponding histograms showing loop area distributions in Figures 7f-i. These loop area distributions indicate that traditional DKL tends to acquire closed hysteresis loops with loop areas close to zero, providing limited information regarding ferroelectric properties. However, spectrum-constrained DKL and structure-constrained DKL sample more opening loops, resulting in deviated loop area distributions and indicating a more informative exploration process. Specifically, the live autonomous microscopy data show that the percentage of sampled opening loops with meaningful ferroelectric characteristics (open loop with a loop area > 0.2) is 17% for the traditional DKL method, 48% for the spectrum constrained DKL method, 57% for the domain wall constrained DKL method, and 22% for the domain constrained DKL method. Additionally, it is worth mentioning that the reason the DKL samples $a$-domains is likely due to high uncertainty in those areas. Predicting spectra from certain locations can be challenging, leading to situations where model uncertainty does not significantly decrease with additional samples because the correlation between structure and property is too low. The constrained methods help bypass these low-correlation areas, which are already known to be problematic, thus enhancing the efficiency of the sampling process.

**Conclusions**

In summary, we developed constrained active learning approaches that incorporate prior knowledge and specific interests for knowledge-informed active learning-driven autonomous experiments. In DKL-driven microscopy, prior knowledge can be applied to either the structural space or the spectral data. These workflows are fully operationalized on autonomous SPM systems and comparative performance of traditional DKL, structure-constrained DKL, and spectrum-constrained is compared. We demonstrated that constrained DKL can lead to more effective exploration in autonomous PFM experiments, via both pre-acquired datasets with known ground truth and operating autonomous PFM.

We note that the developed workflows can be represented as an example from single-step to multiple-step decision making. We expect that this approach can further be extended to build the expert mixture gated workflows, where the role of gate agent is to select the problem-appropriate DKL model, and the latter controls the exploration of image space. These multiple DKL models share the common training data but differ in the possible reward functions and make different exploration decision. This approach can be seamlessly extended to other microscopy and imaging techniques, accelerating scientific discovery. Furthermore, the constraint concept can be adapted for general Bayesian optimization in material discovery across a broad range of autonomous experimental fields.

**Acknowledgements**

This effort (constrained DKL development, PFM measurements) is supported by the Center for Nanophase Materials Sciences (CNMS), which is a US Department of Energy, Office of Science User Facility at Oak Ridge National Laboratory. U.P. and S.V.K. acknowledges support from the Center for Nanophase Materials Sciences (CNMS) user facility which is a U.S. Department of Energy Office of Science User Facility, project no. CNMS2023-B-02196. U.P. and S.V.K. acknowledges support from high performance computing facility, ISAAC and Institute for Advanced Materials & Manufacturing (IAMM) at University of Tennessee Knoxville (UTK). The work was partially supported (SVK) by the Center for Advanced Materials and Manufacturing (CAMM), the NSF MRSEC center. The work was partially supported (UP) AI Tennessee Initiative at University of Tennessee Knoxville (UTK). H.F. acknowledge the support by MEXT Program: Data Creation and Utilization Type Material Research and Development (JPMXP1122683430).

**Authors Contribution**

Y.L. conceived the project, developed the constrained DKL approach, and wrote the manuscript. U.P. and Y.L. performed the analysis and measurement. All authors edited the manuscript.

**Conflict of Interest**

The authors declare no conflict of interest.

**Data Availability**

The constrained DKL approach developed in this work is publicly available at Github https://github.com/yongtaoliu/constrained_DKL.